\documentclass[prd,reprint,nofootinbib,showpacs]{revtex4-1}
\usepackage{amsmath, amssymb,bm,xcolor,graphicx}
\usepackage[pdftitle={Bounds on extra dimensions from micro black holes in the context of the metastable Higgs vacuum},%
    pdfauthor={Katherine J. Mack, Robert McNees},%
    pdfsubject={},%
    pdfkeywords={black holes, ultra high energy cosmic rays, Higgs vacuum, metastability, vacuum decay, extra dimensions},
colorlinks=true,citecolor=blue]{hyperref}
\usepackage[T1]{fontenc}
\usepackage{fontawesome}  
  
\definecolor{plum}{rgb}{0.36078, 0.20784, 0.4}
\definecolor{chameleon}{rgb}{0.30588, 0.60392, 0.023529}
\definecolor{cornflower}{rgb}{0.12549, 0.29020, 0.52941}
\definecolor{twitterblue}{RGB}{64,153,255}
\definecolor{scarlet}{rgb}{0.8, 0, 0}
\definecolor{brick}{rgb}{0.64314, 0, 0}
 
\newcommand{\twitter}[1]{\href{https://twitter.com/#1}{\textcolor{twitterblue}{\faTwitter}\,\tt \textcolor{twitterblue}{@#1}}}
    
\newcommand{\eV}{\text{eV}}

\newcommand{\difffluxunits}{\text{m}^2\cdot \text{s} \cdot \text{sr} \cdot \eV}
\newcommand{\ts}[1]{\text{\tiny{#1}}}
\newcommand{\ms}[1]{\text{\tiny{$#1$}}}

\newcommand{\nts}{\!\!}
\newcommand{\bns}{\!\!\!\!}

\begin{document}

\title{Bounds on extra dimensions from micro black holes in the context of the metastable Higgs vacuum}
\author{Katherine J. Mack}
\email{kmack@ncsu.edu}
\thanks{\twitter{AstroKatie}}
\affiliation{North Carolina State University, Department of Physics, Raleigh, NC 27695-8202, USA}
\author{Robert McNees}
\email{rmcnees@luc.edu}
\thanks{\twitter{mcnees}}
\affiliation{Loyola University Chicago, Department of Physics, Chicago, IL 60660, USA.}

\begin{abstract}
\noindent We estimate the rate at which collisions between ultra-high energy cosmic rays can form small black holes in models with extra dimensions. If recent conjectures about false vacuum decay catalyzed by black hole evaporation apply, the lack of vacuum decay events in our past light cone may place new bounds on the black hole formation rate and thus on the fundamental scale of gravity in these models. For theories with fundamental scale $E_{*}$ above the Higgs instability scale of the Standard Model, we find a lower bound on $E_{*}$ that is within about an order of magnitude of the energy where the cosmic ray spectrum begins to show suppression from the GZK effect. Otherwise, the abundant formation of semiclassical black holes with short lifetimes would likely initiate vacuum decay. Assuming a Higgs instability scale at the low end of the range compatible with experimental data, the excluded range is approximately $10^{17} \,\text{eV} \lesssim E_{*} \leq 10^{18.8}\,\text{eV}$ for theories with $n=1$ extra dimension, narrowing to $10^{17}\,\text{eV} \lesssim E_{*} \leq 10^{18.1}\,\text{eV}$ for $n=6$. These bounds rule out regions of parameter space that are inaccessible to collider experiments, small-scale gravity tests, or estimates of Kaluza-Klein processes in neutron stars and supernovae.\\ 
\end{abstract}

\maketitle

\vskip 1pc

\section{Introduction}
\label{sec:Intro}

In models with extra dimensions, the fundamental scale of gravity may be lower than the four-dimensional Planck scale, $M_\ts{Pl}$. This presents the possibility that high-energy collisions between particles, for instance in colliders or via cosmic rays, may form black holes if a high enough center-of-mass energy is achieved \cite{Banks:1999gd,Giddings:2001bu,Kanti:2004nr,Giddings:2008gr}. Large extra dimensions, if discovered, would constitute new physics and potentially provide an explanation for the relative weakness of gravity in relation to the other fundamental forces. In addition to searches for microscopic black holes formed in particle collisions, experimental constraints on extra dimensions have generally come from searches for modifications of the inverse-square force law of gravity at small scales or from signatures of Kaluza-Klein gravitons or other exotic particles. Constraints on the higher-dimensional fundamental scale depend on the number of extra dimensions proposed, with collider limits in the TeV range and astrophysical limits as high as $\mathcal{O}(10^2)$-$\mathcal{O}(10^3)$ TeV \cite{Kanti:2004nr, Patrignani:2016xqp}.

We present new limits from black hole creation in the context of recent work proposing that Hawking evaporation of black holes can induce the decay of the standard model Higgs vacuum \cite{Gregory:2013hja,Burda:2015yfa,Burda:2016mou,Cuspinera:2018woe}. These papers argue that the nucleation of a bubble of true vacuum in general precedes the final evaporation of the black hole, suggesting that any production of black holes with evaporation times less than the age of the Universe in our past light cone should have already led to vacuum decay. The most recent work in the series \cite{Cuspinera:2018woe} explicitly confirms that evaporating black holes formed in theories with extra dimensions are capable of seeding vacuum decay. The decay of the false vacuum is a dramatic consequence that presents an unmistakable (and fatal) observational signature of microscopic black hole production. Thus, its non-observation allows us to place limits on the higher-dimensional fundamental scale, excluding a range of values that are several orders of magnitude beyond the scales probed by other tests involving micro black hole production, such as via signs of Hawking evaporation in colliders or from nearby cosmic ray collisions.

Our analysis relies on two main assumptions, both of which come with some important caveats that we describe here. The first is the \textit{metastability of the Higgs vacuum}, as implied by recent measurements of the Higgs boson and top quark masses \cite{Buttazzo:2013uya}. This result is based on the validity of the Standard Model of particle physics, so any beyond-Standard-Model (BSM) physics may alter the effective potential of the Higgs field in a way that rescues our vacuum from metastability \cite{Branchina:2014rva}. A high energy scale of inflation, were it to be confirmed, would give evidence that new physics stabilizes the vacuum in some way, since high-energy-scale inflationary fluctuations would likely have instigated a transition to the true vacuum in the early universe \cite{Fairbairn:2014zia, Hook:2014uia}. While we fully recognize (and, in fact, hope) that vacuum metastability ends up being ruled out by BSM physics or a better understanding of inflation, we will, for the purpose of this study, rely on the great successes of the Standard Model to justify the apparent metastability of the Higgs vacuum as an observational tool for establishing constraints on higher-dimensional theories. Second, we are assuming that the results of \cite{Gregory:2013hja, Burda:2015yfa,Burda:2016mou} hold in a qualitatively similar way for theories with more than four spacetime dimensions, i.e., that \textit{black hole evaporation can seed vacuum decay}. This was explored in \cite{Cuspinera:2018woe}, where the authors construct an approximate instanton solution for a braneworld black hole in a theory with one extra dimension and then estimate that their results extend to regimes where small black holes are produced in particle collisions. We will apply these results beyond one extra dimension, though earlier calculations suggest the effect may be somewhat suppressed \cite{Burda:2015yfa}.\footnote{The reduced branching ratio of false vacuum decay rate to the Hawking evaporation rate may be offset by the production of large numbers of black holes.} More importantly, the conclusions of \cite{Cuspinera:2018woe} require the instability scale $\Lambda_{I}$ of the Higgs vacuum to lie below the fundamental scale $E_{*}$ of the higher-dimensional theory. Otherwise, the Standard Model calculation of the Higgs potential no longer applies. For our analysis, we must assume that the instability scale is at the low end of the range consistent with experimental limits. The most likely range calculated by \cite{Buttazzo:2013uya} is $\Lambda_{I} \sim 10^{19}-10^{20}\,\eV$, with some uncertainty around that value. For our analysis to be fully reliable, we require scales no higher than $\Lambda_{I} \sim 10^{18}\,\eV$ for theories with one or two extra dimensions and $\Lambda_{I} \sim 10^{17}\,\eV$ for theories with up to six. This is an important qualifier on our main results, and will be discussed in more detail at the end of the paper.

The structure of our calculation is as follows. Assuming that the Higgs vacuum is metastable and that its decay is catalyzed by black hole evaporation, we take its persistence as evidence against black hole evaporation in our past light cone. While this observation can also constrain the production of low mass primordial black holes in the early universe, we apply it here to the production of microscopic black holes in particle collisions. Specifically, we consider the formation of black holes in collisions between ultra-high-energy (UHE) cosmic rays, in theories with extra dimensions and a fundamental scale well below the four-dimensional Planck scale. If the instability scale for the Higgs vacuum is low enough, this allows us to exclude a range of values for the fundamental scale that are not probed by the accelerator and astrophysical processes associated with current bounds. For a given value of the fundamental scale, we use the UHE cosmic ray spectrum from the Auger experiment (see Section~\ref{sec:Collisions} and ref. \cite{Abreu:2011pj}) to make a conservative estimate of the number of black holes formed in particle collisions in our past light cone. We note that the measured cosmic ray spectrum includes a steep drop-off at high energies. This is believed to be due to the GZK effect \cite{Greisen:1966jv, Zatsepin:1966jv}, which prevents the highest-energy cosmic rays from traveling unimpeded across cosmological distances. If this is the case, then it is likely that even more energetic particles are plentiful in parts of the cosmos where they are accelerated by high-energy astrophysical processes. But without knowing more about the mechanisms involved we restrict our analysis to cosmic rays with energies below the GZK cut-off. Thus, our calculation probably under-estimates both black hole formation rates and the maximum center-of-momentum energies achieved in these collisions. As a result, the excluded range for the fundamental scale of theories with extra dimensions may extend to even higher values than those we present here.

In Section \ref{sec:Collisions} we discuss a method for estimating the number of collisions that have taken place in our past light cone between UHE cosmic rays, and review the Pierre Auger Observatory's spectrum of these particles. In Section \ref{sec:BHFormation} we extend this to collisions capable of forming black holes in higher-dimensional theories, obtain bounds on the fundamental scale of these theories in Section \ref{sec:Bounds}, and then discuss these results in Section \ref{sec:Discussion}. Appendix \ref{app:Semiclassical} considers the various criteria that must be satisfied for a reliable semiclassical analysis of black hole formation, and Appendix \ref{app:AnalyticalN} discusses an analytical result for black hole formation rates that supports the numerical results used in the main text.

\section{Collisions of Ultra-High-Energy Cosmic Rays}
\label{sec:Collisions}

At ultra-high energies, cosmic rays are rare enough that we expect interactions between them to be exceedingly infrequent. But on timescales comparable to the age of the Universe, even a low rate can lead to an appreciable number of collisions with center-of-momentum (CM) energies several orders of magnitude greater than anything that can be achieved in existing accelerators. 
Let us quickly review the estimate of collisions between UHE cosmic rays with energies above $10^{20}\,\eV$ given by Hut and Rees in \cite{Hut:1983xa}.

Assuming a homogeneous and isotropic distribution, the density of UHE cosmic rays with energy greater than $E$ is proportional to the integrated flux 
\begin{gather}\label{density}
	\rho(E) = \frac{4\pi}{c}\,\int_{E}^{\infty} \!\!\! dE' \,J(E') ~,
\end{gather}
where $J(E)$ is the differential flux. For constant density $\rho$ and cross section $\sigma$, the rate of collisions per particle is $\rho  \sigma  c$, and the overall rate of collisions per unit volume is
\begin{gather}\label{RateR}
	R = \rho^2 \sigma c ~.
\end{gather}
The total number of collisions in our past light cone is given by this rate times the spacetime volume
\begin{gather}
	\mathcal{N} = R\,c^{3} T^4 ~,
\end{gather}
where $T$ is the time over which these collisions have occurred and our assumptions hold. Hut and Rees calculated the density of cosmic rays above $10^{20}\,\eV$ using the differential flux given by Cunningham et. al. in \cite{1980ApJ...236L..71C}
\begin{gather}\label{eq:CunninghamFlux}
	J(E) = \frac{1.14\times 10^{-33}}{\difffluxunits} \left(\frac{10^{19}\,\eV}{E}\right)^{2.31} ~.
\end{gather} 
Then \eqref{density} gives a density of $1.8 \times 10^{-23}\,\text{m}^{-3}$. The cross section is taken to be of order the Compton wavelength squared
\begin{gather}\label{ComptonSigma}
	\sigma(E) \simeq \left(\frac{2\pi \hbar c}{E}\right)^2 ~,
\end{gather}
which at $10^{20}\,\eV$ gives $1.5 \times 10^{-52}\,\text{m}^2$. For their order of magnitude estimate, Hut and Rees use $\sigma \simeq 10^{-52}\,\text{m}^{2}$. The per-particle rate of collision is then $\simeq 3 \times 10^{-67}\,\text{s}^{-1}$, and the rate of collisions per unit volume comes out to $R \simeq 3 \times 10^{-90}\,\text{m}^{-3}\,\text{s}^{-1}$. Taking the age of the Universe to be about $T = 10^{10}\,\text{yr}$, the number of collisions in the past light cone is roughly $\mathcal{N} \simeq 8 \times 10^{5}$. Hut and Rees give a final estimate of $\mathcal{N} \approx 10^{5}$. 

\begin{figure} 
	\vskip0.5em  
    \includegraphics[width=\columnwidth{}]{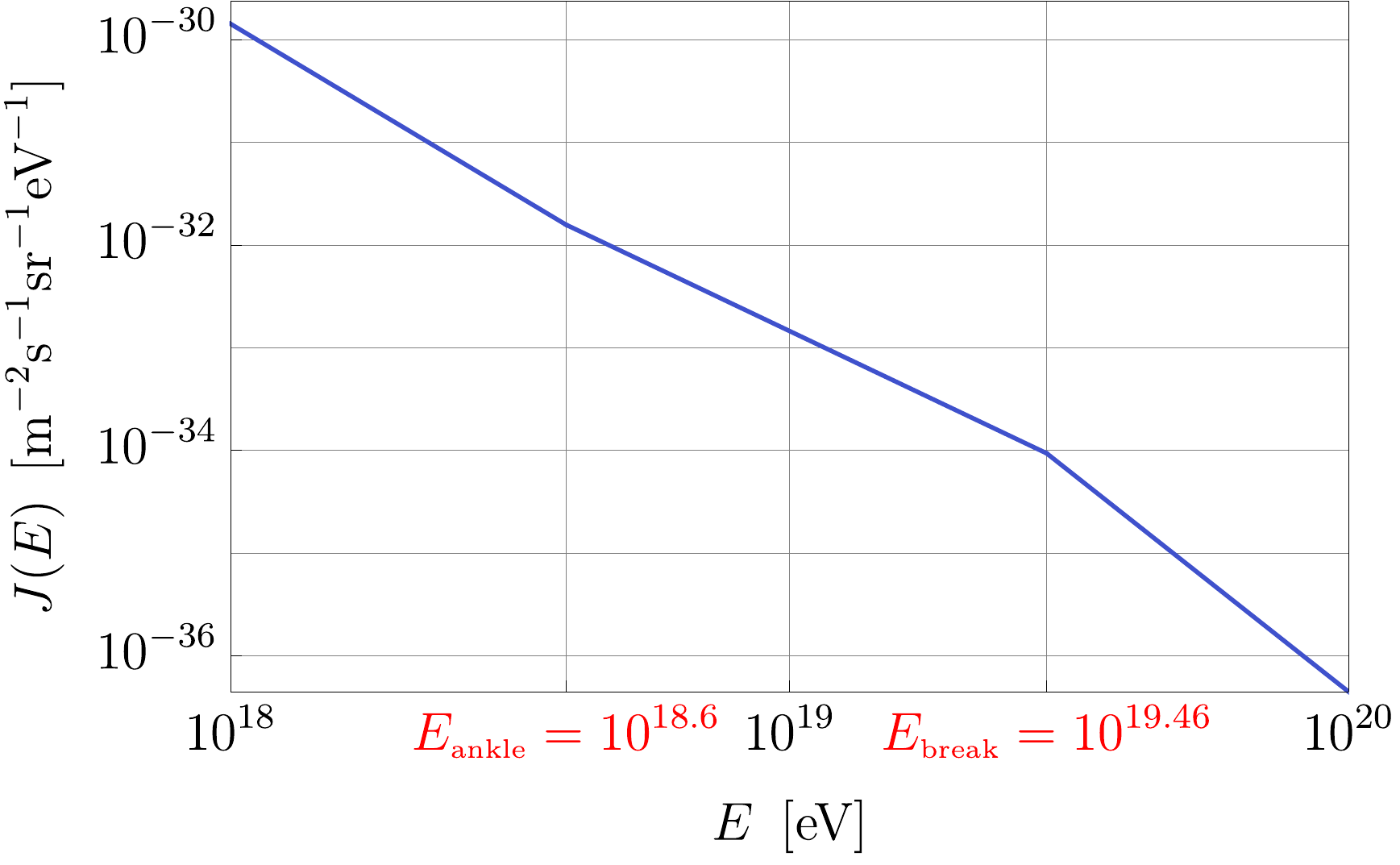}
	\caption{The Auger spectrum of UHE cosmic rays with $E > 10^{18}\,\eV$, approximated as a set of power laws \cite{Abreu:2011pj}.}
	\label{fig:spectrum}
\end{figure}

For our calculations, we will use more recent results for the differential flux of UHE cosmic rays in place of \eqref{eq:CunninghamFlux}. The Pierre Auger Observatory is a hybrid cosmic-ray observatory consisting of surface Cerenkov detectors and air-shower observing telescopes, which allows it to collect large samples of cosmic rays across a wide range of energies. Its measurements of the cosmic ray energy spectrum above $10^{18}\,\eV$ are well described by a series of power laws with free breaks between them, or else by broken power laws with an additional smooth suppression factor at the highest energies \cite{Abraham:2010mj, Abreu:2011pj, Aab:2015bza, Aab:2017njo}. Here we adopt the values given in \cite{Abreu:2011pj}, with the differential flux in each range of energies taking the form
\begin{gather}
	J(E) \propto E^{-\gamma} ~.
\end{gather}
The flux is shown in Fig.~\ref{fig:spectrum}. Below the `ankle' energy, $E_\text{\tiny{ankle}} =10^{18.61\pm0.01}\,\eV$, the spectral index is $\gamma_{1} = 3.27 \pm 0.02$. Above the ankle energy, but below the `break' energy $E_\text{\tiny{break}} = 10^{19.46\pm0.03}\,\eV$, the spectral index flattens to $\gamma_2 = 2.59\pm0.02$. Above the break energy the spectral index drops off to $\gamma_3 = 4.3\pm0.2$. The spectrum between $E_\ts{ankle}$ and $E_\ts{break}$ is thought to possibly represent the transition to a population of extragalactic cosmic rays, while the steep fall off above $E_\ts{break}$ is likely due to the GZK effect \cite{Greisen:1966jv, Zatsepin:1966jv}.

Repeating the estimate of Hut and Rees with the Auger spectrum yields fewer collisions above $10^{20}$ eV -- on the order of a few thousand -- because of the steep drop off in the flux above $E_\ts{break}$ that is not accounted for in \eqref{eq:CunninghamFlux}. Indeed, the presence of this feature in the spectrum suggests that the upper limit in \eqref{density} should not extend to arbitrarily high energies. The GZK effect prevents cosmic rays from traveling cosmological distances with energies greater than $E_\ts{break}$.\footnote{It is possible that the GZK effect is only partly responsible for the drop-off in flux above $E_\ts{break}$, which may also reflect, for instance, the maximum energies that can be achieved by the sources that accelerate the particles \cite{Abraham:2010mj}. While its origin does not impact our estimates, we interpret $E_\ts{break}$ as the scale associated with the GZK effect.} So it seems unwarranted to assume a homogeneous and isotropic distribution for cosmic rays at those energies over the full volume of our past lightcone. Without knowing more about the origin of UHE cosmic rays, we will conservatively limit all of our calculations to cosmic rays with energies below the break energy in the Auger spectrum.

The approximation \eqref{RateR} for the rate of collisions per unit volume treats all particles as if they had roughly the same energy, with a constant cross section for collisions. We can refine the estimate by dropping these assumptions, accounting instead for all collisions above a given CM energy and including the energy dependence of the cross section. Assuming once again a homogeneous and isotropic distribution of UHE cosmic rays, the rate per unit volume of collisions with CM energy greater than $E$ is given by
\begin{multline}\label{RefinedRate}
	R = \frac{16\pi^{2}}{c}\,\int_{E_\ts{min}}^{E_\ts{break}} \bns\bns\nts\nts dE' dE'' \int_{0}^{1}\nts du\,\sigma(E_\ts{CM}) \\
	J(E') J(E'') \,\Theta(E_\ts{CM} - E)~,
\end{multline}
where $u = (1-\cos\psi)/2$ is related to the angle $\psi$ between the momenta of the colliding particles, $E_\ts{CM} = 2\sqrt{E'E''u}$ is the CM energy, and the Heaviside theta function restricts the domain of integration to collisions with $E_\ts{CM} > E$. The upper limit in the energy integrals is taken to be $E_\ts{break}$, which restricts the calculation to cosmic rays with energies below the GZK cut-off, while the lower limit is
\begin{gather}
	E_\ts{min} = \frac{E^2}{4\,E_\ts{break}}~,
\end{gather}
which is the minimum energy of a particle that can participate in a collision with CM energy of at least $E$. As before, the number of events in our past lightcone is $\mathcal{N} = R\,c^{3} T^{4}$. In our calculations, we will take $T = 10^{10}$ years. This may be a conservative assumption, as the production rate for UHE cosmic rays is likely to have been higher toward the early part of that time window, closer to the peak of active galactic nucleus (AGN) activity around a redshift of 2.

In the next section we will employ \eqref{RefinedRate} to estimate the rate of black hole formation in models where the fundamental scale of gravity is below the maximum CM energies accessible in collisions between UHE cosmic rays.

\section{Black Hole Formation via Cosmic Ray Collision}
\label{sec:BHFormation}

In higher-dimensional theories the fundamental scale of gravity may be lower than $M_\ts{Pl}$. We will consider a generic $4+n$-dimensional theory with fundamental scale $M_{*}=E_{*}/c^2$ related to the Newton's constant by
\begin{gather}\label{GM*}
	G_{N}^{\,\ts{(4+n)}} = \frac{c^{n+5}\,\hbar^{n+1}}{E_{*}{}^{2+n}} ~.
\end{gather}
If the scale $E_{*}$ is low enough, collisions between UHE cosmic rays at sufficiently high CM energy are expected to form black holes of mass $M_\ts{BH} \sim E_\ts{CM}/c^{2}$.

For our estimates of black hole formation via scattering to make sense, the black holes should be large enough that a semiclassical treatment is appropriate. We enforce this by considering only black holes with entropy above some minimum value $S_\ts{min}$ (see Appendix \ref{app:Semiclassical} for a brief discussion). This implies that the ratio of $M_\ts{BH}/M_{*}$ must be greater than 
\begin{gather}\label{eq:lambda}
	\lambda = \frac{n+2}{4\pi}\,\left(\frac{\pi^{\frac{n+3}{2}}}{2\,\Gamma(\frac{n+3}{2})}\right)^{\frac{1}{n+2}} \big(S_\ts{min}\big)^{\frac{n+1}{n+2}} ~.
\end{gather}
We will typically take $S_\ts{min} = 10^{2}$, which implies that $M_\ts{BH}$ must be greater than $M_{*}$ by a factor that is $\mathcal{O}(10)$ for $n=1$ and increases to $\mathcal{O}(10^2)$ for $n=6$. Neglecting energy loss during the formation process, $M_\ts{BH} = E_\ts{CM}/c^2$ and collisions with $E_\ts{CM} \geq \lambda\,E_{*}$ are considered to form semiclassical black holes.

We also require that the black holes be small enough compared with the compactification scale $L$ that a flat-space approximation is valid. For a discussion of this requirement, see Appendix \ref{app:Semiclassical}.

For the collisions between UHE cosmic rays in the previous section, the cross section \eqref{ComptonSigma} was proportional to the square of the Compton wavelength and hence decreased at higher energies. But at CM energies well above $E_{*}$ the cross section for black hole formation exhibits the opposite behavior. The geometric cross section for black hole formation is \cite{Argyres:1998qn}-\cite{Kaloper:2007pb}
\begin{gather}\label{eq:BHCS}
	\sigma_\ts{BH} = \mathcal{O}(1)\,\pi r_\ts{H}^{2} ~,
\end{gather}
where $r_\ts{H}$ is the horizon radius of a black hole of mass $M_\ts{BH} = E_\ts{CM}/c^2$, and an overall factor of order 1 reflects various corrections. Collisions at higher energies produce black holes with larger mass, and hence larger horizon radius, resulting in a cross section that grows as a positive power of $E_\ts{CM}$.

Assuming the collision forms a Schwarzschild black hole, the horizon radius in $4+n$ dimensions is \cite{Myers:1986un}
\begin{gather}\label{HorizonRadius}
	r_\ts{H} = \frac{\hbar\,c}{E_{*}}\,\left(\frac{M_\ts{BH} c^2}{E_{*}}\right)^{\frac{1}{n+1}} \,\left(\frac{8\pi}{n+2}\,\frac{\Gamma(\tfrac{3+n}{2})}{\pi^{\frac{3+n}{2}}}\right)^{\frac{1}{n+1}}~.
\end{gather}
Then, up to the $\mathcal{O}(1)$ factor in Eq.~\eqref{eq:BHCS}, the cross section for a collision forming a black hole with $M_\ts{BH} = E_\ts{CM}/c^{2} $ is
\begin{gather}\label{eq:BHCS2}
	\sigma_\ts{BH}^\ts{(4+n)} =  \left(\frac{\hbar\,c}{E_{*}}\right)^2\,\left(\frac{E_\ts{CM}}{E_{*}}\right)^{\frac{2}{n+1}} \,\left(\frac{8 \,\Gamma(\tfrac{3+n}{2})}{n+2}\right)^{\frac{2}{n+1}} ~.
\end{gather} 
Using this cross section in Eq.~\eqref{RefinedRate}, we can estimate the rate at which black holes are formed by collisions between UHE cosmic rays in a higher-dimensional theory with fundamental scale $E_{*}$.

As an example, consider a theory with one extra dimension ($n=1$). Using Eq.~\eqref{eq:lambda}, black holes with entropy $S_\ts{BH} \geq 10^{2}$ have mass $M_\ts{BH} \geq 8.76\,M_{*}$. The number of such black holes formed in our past light cone by collisions between UHE cosmic rays is approximately
\begin{widetext}
\begin{gather}\label{BHExample}
	\mathcal{N} = c^{3}\,T^{4} \frac{128\pi^{2}}{3\,c}\,\left(\frac{\hbar\,c}{E_{*}}\right)^{2}\int_{E_\ts{min}}^{E_\ts{break}} \bns\bns\nts dE' dE'' \int_{0}^{1}\nts du\,\frac{\sqrt{E'\,E''\,u}}{E_{*}} J(E') J(E'') \,\Theta\big(2\sqrt{E'\,E''\,u} - 8.76\,E_{*}\big)~.
\end{gather}	
\end{widetext}
Fixing a fundamental scale allows us to evaluate this expression explicitly to determine a number of black hole-producing events above that scale. As an illustrative calculation, we use the Auger results for the differential flux, $T = 10^{10}\,\text{yr}$, and a fundamental scale of $10^{18.5}\,\eV$. The numerical evaluation of this integral gives $\mathcal{N} \simeq 1.6 \times 10^{11}$. Raising the fundamental scale to $10^{18.8}\,\eV$ lowers the number of events to $\mathcal{N} \simeq 2.6 \times 10^{6}$.

Since we restrict our attention to UHE cosmic rays with energy below the GZK cut-off at $E_\text{\tiny{break}} = 10^{19.46}\,\eV$, the maximum possible CM energy in our analysis is $2\,E_\ts{break} = 10^{19.76}\,\text{eV}$. This places an upper limit on the mass of the black hole, so the requirement $M_\ts{BH} \geq \lambda\,M_{*}$ implies that our analysis is valid only for $E_{*} < 2\,E_\ts{break}/\lambda$. In this example with $n=1$, the largest fundamental scale we can consider is $10^{18.82}\,\text{eV}$, and $\mathcal{N}$ plunges to zero as $E_{*}$ approaches this value. The drop-off is steep enough that the difference between $E_{*}$ for $\mathcal{N} \sim 10^{6}$ and $\mathcal{N} \sim 10^{3}$ is much less than the uncertainties in $E_\ts{break}$ and other factors. If vacuum decay triggered by black hole evaporation is as likely as the claims of \cite{Gregory:2013hja, Burda:2015yfa, Burda:2016mou, Cuspinera:2018woe}, then essentially any value of $E_{*}$ up to $2\,E_\ts{break}/\lambda$ results in too many black holes being formed.

Note that by raising the maximum cosmic ray energy in Eq.~\eqref{RefinedRate}, $E_{*}$ could be greater than $2\,E_\ts{break}/\lambda$ by a factor of $\sim$5 and still allow a significant number of black holes to form and evaporate over the history of our universe. But the majority of those collisions would involve particles with energies above the GZK cut-off, and as explained in the previous section, our assumption of a homogeneous and isotropic distribution seems questionable for that population of UHE cosmic rays.\\

\section{Bounds on $E_{*}$ from Black Hole Formation}
\label{sec:Bounds}

Now we calculate the number of black holes formed via collisions between UHE cosmic rays for different numbers of extra dimensions $n$, and use this to establish a range of excluded values for the fundamental scale $E_{*}$.

\begin{figure}
    \includegraphics[width=\columnwidth{}]{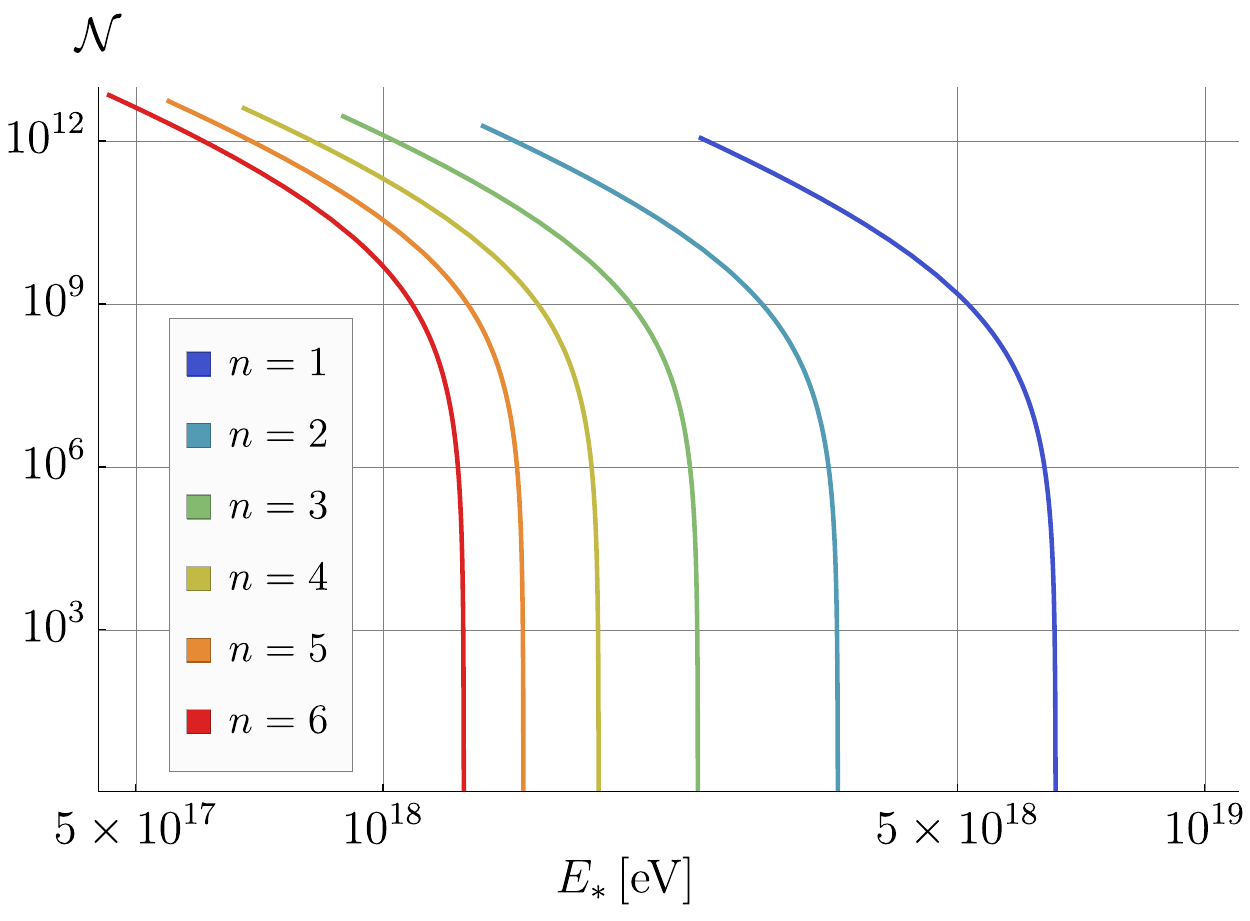}
	\caption{Number of collisions $\mathcal{N}$ forming a black hole with entropy $S_\ts{BH}\geq 10^2$ over the past $T = 10^{10}\,\text{yr}$, as a function of the fundamental scale. The colored lines are for different values of $n$, the number of extra dimensions. $\mathcal{N}$ drops off rapidly as $E_{*}$ approaches $2E_\ts{break}/\lambda$.}
	\label{fig:NvsEfund}
\end{figure}

As in the previous section, we consider only black holes with entropy above a minimum value $S_\ts{min}$ that justifies the use of semiclassical methods. This implies $M_\ts{BH} \geq \lambda\,M_{*}$, where $\lambda$ is given by Eq.~\eqref{eq:lambda}. Then the number of black holes formed is approximately
\begin{widetext} 
\begin{gather}\label{GeneralN}
	\mathcal{N} = c^{3}\,T^{4}\,\frac{16\pi^{2}}{c}\,\int_{E_\ts{min}}^{E_\ts{break}} \bns\bns\nts\nts dE' dE'' \int_{0}^{1}\nts du\,\sigma_\ts{BH}^{\ms{(4+n)}}(E_\ts{CM}) J(E') J(E'') \,\Theta\big(E_\ts{CM} - \lambda\,E_{*}\big)~,
\end{gather}	
\end{widetext}
where $E_\ts{CM} = 2\,\sqrt{E'E''u}$. Since we only consider UHE cosmic rays with energies below $E_\ts{break}$, these collisions can only form semiclassical black holes when the fundamental scale satisfies $E_{*} \leq 2E_\ts{break}/\lambda$. Because the Auger spectrum is defined piecewise for different values of the energy, Eq.~\eqref{GeneralN} is most easily evaluated numerically. The number of black holes formed as a function of $E_{*}$ is shown in Fig.~\ref{fig:NvsEfund}, for $1 \leq n \leq 6$.  

As in the $n=1$ example, $\mathcal{N} \gg 1$ for all values of $E_{*}$ up to the maximum value $2E_\ts{break}/\lambda$ that can be probed using our method. For these values of $E_{*}$, collisions between UHE cosmic rays that form rapidly evaporating black holes are plentiful within our past light cone. Thus, avoiding vacuum decay catalyzed by the evaporation of these black holes excludes fundamental scales in the range
\begin{gather}
	\Lambda_{I} < E_{*} \leq \frac{2\,E_\ts{break}}{\lambda} = \frac{10^{19.76}\,\text{eV}}{\lambda}~,
\end{gather} 
where $\lambda$ is given in Eq.~\eqref{eq:lambda}. Collisions between cosmic rays at even higher energies are of course possible, and would raise the upper end of the range of excluded values for $E_{*}$, but they cannot be reliably estimated with the method used here.
  
\begin{figure} 
    \includegraphics[width=\columnwidth{}]{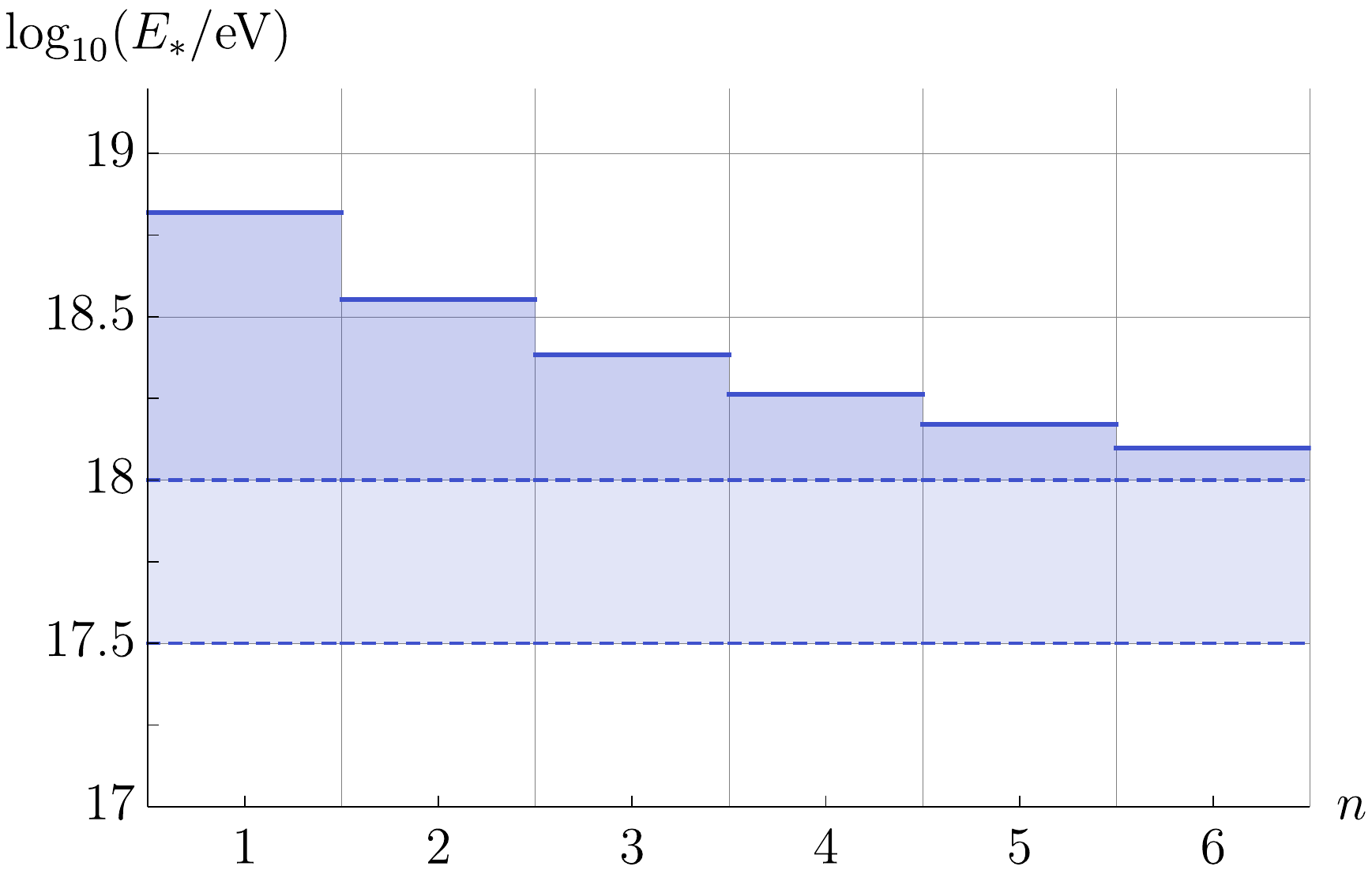}
	\caption{Excluded range for the fundamental scale for $1 \leq n \leq 6$ with $S_\ts{BH} \geq 10^{2}$ and $\Lambda_{I} \sim 10^{17}\text{eV}$, where $n$ is the number of extra dimensions, $S_\ts{BH}$ is the entropy of the black hole (set to ensure the validity of a semiclassical treatment), and $\Lambda_{I}$ is the Higgs instability scale. Existing constraints are at scales $\mathcal{O}(10^2)$-$\mathcal{O}(10^3)$ TeV, well below the plot range shown here. The darker shaded region of the plot is reliably excluded by our constraints. The lighter shaded region between $10^{17.5}\,\text{eV}$ and $10^{18}\,\text{eV}$, where $E_{*}$ is comparable to $\Lambda_{I}$, indicates uncertainties surrounding the applicability of our calculation.   
} 
	\label{fig:LowerBound}
\end{figure}

\begin{figure} 
    \includegraphics[width=\columnwidth{}]{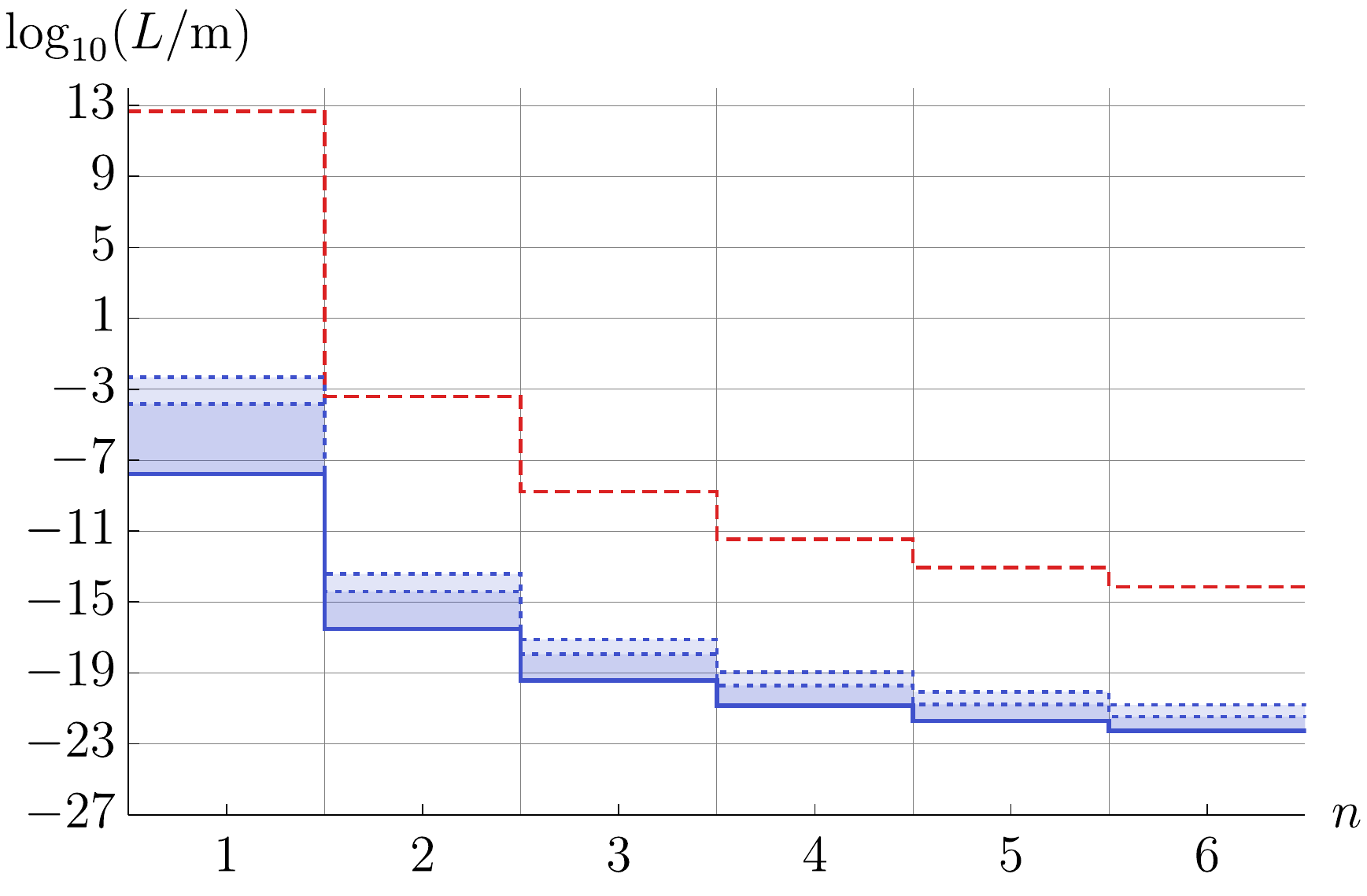}
	\caption{Bounds on the size of extra dimensions (blue), with typical values for TeV-scale gravity (red), for different values of the number of extra dimensions, $n$. The blue shaded region of the plot is excluded by our constraints.}
	\label{fig:Lplot}
\end{figure}

The excluded range of $E_{*}$ for different values of $n$ is shown in Fig.~\ref{fig:LowerBound}, assuming a Higgs instability scale of $\Lambda_{I} \sim 10^{17}\,\text{eV}$. 
For $S_\ts{BH} \geq 10^{2}$, the upper end of this range is more or less within an order of magnitude of the Auger break energy, and safely above our assumed instability scale for the Higgs vacuum. A more conservative condition $S_\ts{BH} \geq 10^{3}$ would require the formation of larger black holes, which may not be consistent with the assumed instability scale. Fig.~\ref{fig:Lplot} translates the excluded range for $E_{*}$ (with  $S_\ts{BH} \geq 10^2$) into a bound on the size of the extra dimensions \eqref{eq:SizeOfL} with the typical size of extra dimensions for TeV-scale gravity \cite{Agashe:2014kda} included for comparison.  
The values for $E_{*}$ and $L$ are summarized in Table \ref{tab:limits}.
\begin{table}[h] 
\def\arraystretch{1.5}	
\begin{tabular}{cccc} \hline \hline
	\,\,$n$\,\, & $\,\,\lambda \,(S_\ts{min} = 10^{2})\,\,$ & $\,\,\log_{10}(E_{*}/\text{eV})\,\,\,$ & $\,\,\log_{10}(L/\text{m})\,\,\,\,\,$ \\ \hline \hline
	$1$ & $8.8 $ & ${18.8}$ & ${-7.8}$ \\
	$2$ & $16.1$ & ${18.6}$ & ${-16.5}$\\
	$3$ & $23.9$ & ${18.4}$ & ${-19.4}$\\
	$4$ & $31.5$ & ${18.3}$ & ${-20.9}$\\
	$5$ & $38.9$ & ${18.2}$ & ${-21.7}$\\
	$6$ & $46.0$ & ${18.1}$ & ${-22.3}$\\ \hline \hline
\end{tabular}
\caption{
Upper end of the excluded range of values for the fundamental scale $E_{*}$, with the associated size $L$ of extra dimensions assuming a toroidal compactification.
}
\label{tab:limits}
\end{table}

The values of $\mathcal{N}$ used in Fig.~\ref{fig:NvsEfund} were calculated by numerically evaluating Eq.~\eqref{GeneralN}. However, it is easy to show that once $E_{*}$ is larger than about $0.8\,E_\ts{break}/\lambda$, both of the UHE cosmic rays participating in the collision must have $E > E_\ts{ankle}$. In that case, the relevant part of the Auger differential flux is given by a single power law and $\mathcal{N}$ can be evaluated analytically as a function of $E_{*}$ and $n$. This is described in more detail in Appendix \ref{app:AnalyticalN}.

\section{Discussion}
\label{sec:Discussion}

We have presented constraints on the fundamental scale and the size of extra dimensions in higher-dimensional theories, based on the non-observation of vacuum decay catalyzed by microscopic black holes. This scenario is based on the mechanism outlined in \cite{Gregory:2013hja, Burda:2015yfa, Burda:2016mou, Cuspinera:2018woe}, in which black holes seed vacuum decay before their evaporation is complete. It assumes the meta-stability of the Higgs vacuum, supported by recent measurements of the mass of the Higgs boson and top quark, at a scale below the fundamental scale of the higher-dimensional theory. While this concept has been used to place limits on the production of primordial black holes \cite{Burda:2015yfa, Burda:2016mou}, ours is the first analysis to place quantitative limits on the fundamental scale of extra dimensional theories based on this method.

Table \ref{tab:otherlimits} summarizes current limits on the fundamental scale and/or size of extra dimensions from a range of methods. Comparing with Table \ref{tab:limits}, limits from vacuum decay catalyzed by black hole evaporation exclude ranges of these parameters that are many orders of magnitude beyond what can be probed with current tests.
\begin{table}
\def\arraystretch{1.5}	
\begin{tabular}{ccccc} \hline \hline
	\,\,Method\,\, & Reference\,\, & \,\,$n$\,\, & $\,\,\log_{10}(E_{*}/\text{eV})\,\,\,$ & $\,\,\log_{10}(L/\text{m})\,\,\,\,\,$ \\
	\hline \hline
	Grav force & \cite{Murata:2014nra} & $2$ & ${12.5}$ & ${-4.36}$ \\
	\hline
	SN1987A & \cite{Hanhart:2001fx} & $2$ & ${13.4}$ & ${-6.18}$\\
	 &  & $3$ & ${12.4}$ & ${-9.10}$\\
	\hline
	NS cooling & \cite{Hannestad:2003yd} & $1$ & ${}$ & ${-4.35}$\\
	 &  & $2$ & ${}$ & ${-9.81}$\\
	 &  & $3$ & ${}$ & ${-11.6}$\\
	 &  & $4$ & ${}$ & ${-12.5}$\\
	 &  & $5$ & ${}$ & ${-13.0}$\\
	 &  & $6$ & ${}$ & ${-13.4}$\\
	\hline
	CMS & \cite{Sirunyan:2017jix} & $2$ & ${13.0}$ & ${}$\\
	 &  & $3$ & ${12.9}$ & ${}$\\
	 &  & $4$ & ${12.8}$ & ${}$\\
	 &  & $5$ & ${12.8}$ & ${}$\\
	 &  & $6$ & ${12.7}$ & ${}$\\
	\hline \hline
\end{tabular}
\caption{Current bounds on extra dimensions from: gravitational force law tests \cite{Murata:2014nra}; constraints on the production of Kaluza-Klein gravitons from the supernova 1987A \cite{Hanhart:2001fx}; constraints based on the expectation that Kaluza-Klein gravitons would decay into photons and heat neutron stars  \cite{Hannestad:2003yd}; and collider searches, the most stringent of which currently provided by the CMS collaboration \cite{Sirunyan:2017jix}. We provide values for both $E_{*}$ and $L$ when provided in the cited references. In other cases, it is possible to deduce the corresponding value via equation \ref{eq:SizeOfL} for the toroidal compactifications considered here. It is of note that the most stringent constraints (SN1987A and NS cooling) require some assumptions about Kaluza-Klein gravitons.}
\label{tab:otherlimits}
\end{table}

Our analysis is limited by the assumptions underlying the conclusions of \cite{Gregory:2013hja, Burda:2015yfa, Burda:2016mou, Cuspinera:2018woe}. The most important of these assumptions is the requirement that $\Lambda_{I} < E_{*}$. If the Higgs instability scale approaches $2\,E_\ts{break}/\lambda$ then the excluded range collapses and there is no bound. Likewise, we are not able to draw any conclusions from this analysis concerning scenarios where $E_{*}$ is below $\Lambda_{I}$.\,\footnote{Recall that the original motivation for large extra dimensions -- a natural explanation for the apparent weakness of gravity -- would require a fundamental scale well below the Higgs instability scale.} The authors of \cite{Cuspinera:2018woe} quote $\Lambda_{I} \sim 10^{17}\,\eV$ as the lowest value consistent with experimental limits on the top quark mass, which is well below our lower bounds on $E_{*}$ for $1 \leq n \leq 6$. They estimate that for $n=1$ extra dimension, with $\Lambda_{I} \sim 10^{17}\,\eV$ and $E_{*} \sim 10^{18}\,\eV$, vacuum decay would be caused by black holes with $M_\ts{BH} \sim 10^{20}\,\eV$. For $E_{*} < 10^{18.8}\,\eV$, we find that a significant number of black holes with $M_\ts{BH} \sim 10^{20}\,\eV$ are produced. Even if the instability scale is as high as $\Lambda_{I} \sim 10^{18}\,\eV$, our $n=1$ (and possibly $n=2$) value for $E_{*}$  seems high enough to justify concerns about vacuum decay. However, the most likely range of values for the instability scale $\Lambda_{I}$ appears to be around $10^{19}\text{-}10^{20}\,\eV$. In that case the fundamental scale $E_{*}$ would have to be greater than $10^{20}$-$10^{21}\,\eV$, which is outside the regime that can be probed with this conservative calculation. On the other hand, UHE cosmic ray observatories have detected particles with energies as high as $E = 3\times 10^{20}\,\text{eV}$. The propagation of such particles on cosmological scales is suppressed by the GZK effect, so our method for estimating the number of collisions forming black holes is not applicable. But if collisions between particles at these energies occur in regions where UHE cosmic rays are produced, then CM energies $E_\ts{CM} \sim 10^{21}$ may be achieved. In that case rapidly evaporating black holes may have been formed for fundamental scales as high as $E_{*} \sim 10^{20}\,\eV$, potentially inducing vacuum decay even if the instability scale is $\Lambda_{I} \sim 10^{19}\,\eV$, which is in the most likely range of values \cite{Buttazzo:2013uya}. For larger values of the instability scale it seems unlikely that black hole formation via UHE cosmic ray collisions could be used to constrain $E_{*}$.

The approach we have taken here comes with important caveats described above and in the introduction. Nevertheless, the possibility of establishing bounds on extra dimensions at scales that are not probed by other methods makes this is a promising direction for continued research. Additionally, our method relies on qualitatively different physics than the tests responsible for the strongest existing constraints on extra dimensions. It does not rely on assumptions about gravitons or other BSM particle physics, and is therefore an interesting complement to existing methods.

We expect that our analysis can be made more robust with improved inferences about the instability scale from collider data, along with a more complete accounting of the full range of high-energy particle interactions in our past light cone. In particular, collisions in regions where UHE cosmic rays are accelerated to energies above the GZK cut-off may achieve higher CM energies than we considered. Such collisions could form black holes for even larger values of the fundamental scale $E_{*}$, making our analysis relevant for a wider range of values of the Higgs instability scale.

\section{Acknowledgments}
RM was supported in part by the National Science Foundation under Grant No.~NSF PHY-1748958 through the KITP Scholars program, and by Loyola University Chicago through a Summer Research Stipend. KJM and RM acknowledge the hospitality of Caltech and the Burke Institute, especially for the workshop ``Unifying Tests of General Relativity'' (July 2016) during which this project was conceived. KJM also acknowledges the Australian Research Council and Melbourne University for support during early stages of this project, and RM thanks the Kavli Institute for Theoretical Physics and the KITP Scholars Program for hospitality and a productive work environment. We are grateful to Leo Stein and Walter Tangarife for helpful comments on an earlier draft of this paper, and to Nima Arkani-Hamed for an important correction that improved the quality of our analysis.

\appendix

\section{Black Hole Formation and Semiclassical Methods}
\label{app:Semiclassical}

Our analysis is based on the formation of black holes in collisions between UHE cosmic rays, which rapidly evaporate via Hawking radiation. Thus, we must consider three questions. First, under what conditions can we say that a collision has formed a black hole? Second, when can those black holes be described semiclassically? And third, since the extra dimensions of spacetime are assumed to be small and compact, when can the black holes be described using results that assume an asymptotically flat space time? 
 
A basic criteria for saying that a black hole has formed is that the decay time should be very long compared to the time scale associated with the formation process. For black holes formed via collision, we take that to mean that the decay time should be much longer than the time needed for the particles to cross a region of linear size $r_\ts{H}$. In $4+n$ dimensions the decay time is of order
\begin{gather}
	\tau_\ts{D} \sim \frac{\hbar}{M_{*} c^{2}}\,\left(\frac{M_\ts{BH}}{M_{*}}\right)^\frac{n+3}{n+1} ~,
\end{gather}
while the crossing time $\tau_\ts{C} \simeq r_\ts{H}/c$ is
\begin{gather}
	\tau_\ts{C} = \frac{\hbar}{M_{*} c^{2}} \frac{1}{\sqrt{\pi}} \left(\frac{8\, \Gamma(\tfrac{3+n}{2})}{n+2}\right)^\frac{1}{n+1} \left(\frac{M_\ts{BH}}{M_{*}}\right)^\frac{1}{n+1} ~.
\end{gather}
In all cases of interest, the $n$-dependent factors in $\tau_\ts{C}$ are $\mathcal{O}(1)$, so the condition $\tau_\ts{D} \gg \tau_\ts{C}$ is equivalent to 
\begin{gather}
	\left(\frac{M_\ts{BH}}{M_{*}}\right)^\frac{n+2}{n+1} \gg 1 ~.
\end{gather} 
The power on the left-hand side of this inequality is always greater than 1, so black holes with $M_\ts{BH} \gg M_{*}$ satisfy $\tau_\ts{D} \gg \tau_\ts{C}$.

To justify a semiclassical treatment, the entropy of the black hole should satisfy $S_\ts{BH} \gg 1$. In any dimension the entropy is given by one-quarter of the horizon area in units of the fundamental length scale. For a non-rotating black hole this is: \begin{gather}
	S_\ts{BH} = \frac{A_\ts{H} c^3}{4\,\hbar\,G_\ts{N}^\ts{(4+n)}} = \frac{\omega_{2+n}\, r_\ts{H}^{2+n} c^{3}}{4\,\hbar\,G_\ts{N}^\ts{(4+n)}} ~,
\end{gather}
where $\omega_{2+n} = 2\pi^{\frac{3+n}{2}}/\Gamma(\tfrac{3+n}{2})$ is the area of a unit $2+n$-sphere. Using Eq.~\eqref{GM*} and Eq.~\eqref{HorizonRadius}, the entropy can be expressed as
\begin{gather}\label{BHEntropy}
	S_\ts{BH} = \left(\frac{4\pi}{n+2}\right)^{\frac{n+2}{n+1}}\left(\frac{4}{\omega_{n+2}}\right)^{\frac{1}{n+1}} \left(\frac{M_\ts{BH}}{M_{*}}\right)^\frac{n+2}{n+1} ~.
\end{gather}
The first two factors give a number greater than 1 for $1 \leq n \leq 9$, and of $\mathcal{O}(1)$ out to $n \sim 35$. So the condition $S_\ts{BH} \gg 1$ is essentially the same as the previous condition, $\tau_\ts{D} \gg \tau_\ts{C}$, in all cases of interest.

For a black hole of mass $M_\ts{BH} = 10\,M_{*}$, the entropy ranges from $S_\ts{BH} \simeq 120$ when $n=1$, down to $S_\ts{BH} \simeq 20$ for $n=6$. 
\begin{figure}
	\vskip0.5em  
    \includegraphics[width=\columnwidth{}]{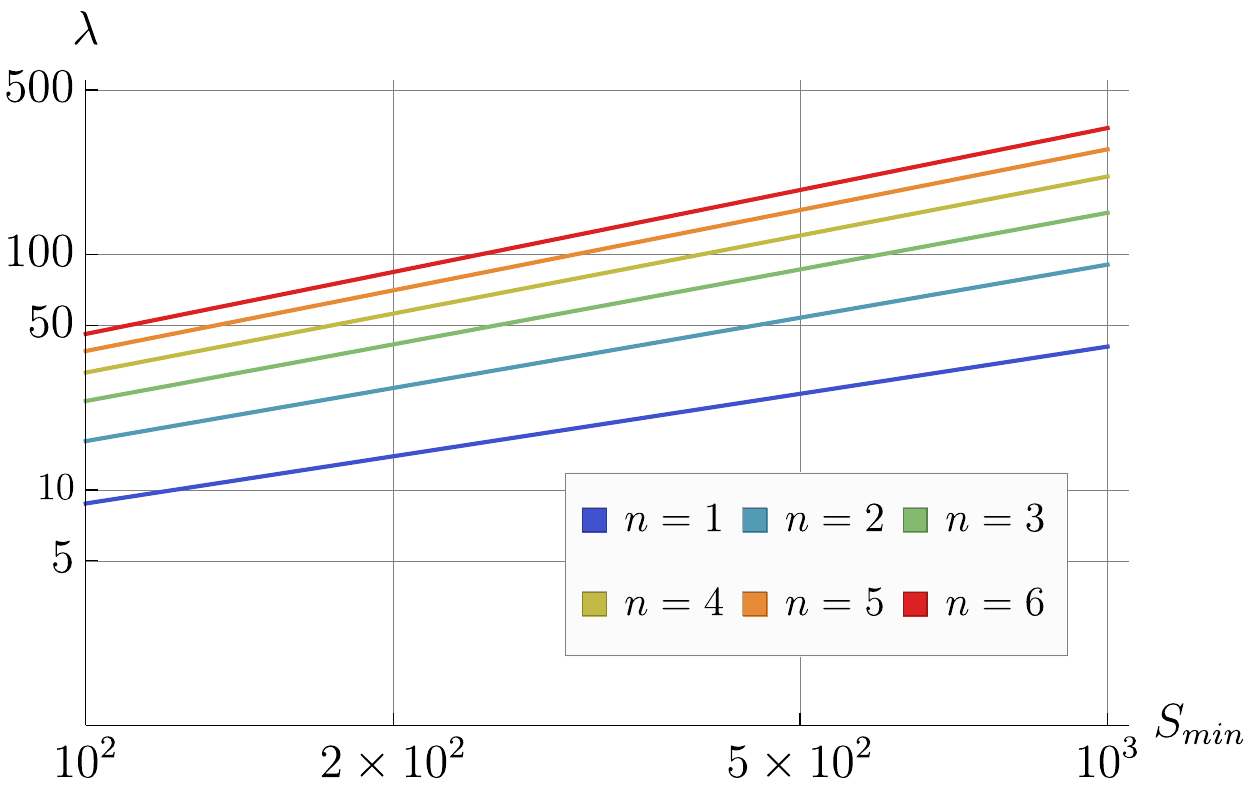}
	\caption{The ratio $\lambda = M_\ts{BH}/M_{*}$ for a black hole with entropy $S_\ts{min}$.}
	\label{fig:lambdaplot}
\end{figure}
Since $S_\ts{BH}$ decreases with $n$ for fixed $M_\ts{BH}/M_{*}$, we will always set a minimum entropy $S_\ts{min}$ that is sufficient to justify semiclassical calculations, and then restrict our attention to black holes with entropy at or above this cut-off. Using Eq.~\eqref{BHEntropy}, this fixes the minimum value $\lambda$ of the ratio $M_\ts{BH}/M_{*}$ for a semiclassical black hole in our analysis as	
\begin{gather}
	\lambda = \frac{n+2}{4\pi}\,\left(\frac{w_{n+2}}{4}\right)^{\frac{1}{n+2}}\left(S_\ts{min}\right)^{\frac{n+1}{n+2}} ~.
\end{gather}
This is shown for $1 \leq n \leq 6$ in Fig.~\ref{fig:lambdaplot}.
Black holes with $M_\ts{BH} \geq \lambda\, M_{*}$ have $S_\ts{BH} \geq S_\ts{min}$.

For the higher dimensional theories considered in this paper we assume that $n$ dimensions are compactified with length scale $L.$ The formulas above assume that spacetime is asymptotically flat, but we may regard them as approximately true when the black hole radius \eqref{HorizonRadius} is much smaller than the compactification scale: $r_\ts{H} \ll L$. In the case of toroidal extra dimensions, the four-dimensional Planck mass is related to the compactification scale of the higher dimensional theory by
\begin{gather}\label{eq:SizeOfL}
	M_\ts{Pl}^{\,2} = (2\pi L)^{n}\,(M_{*})^{2+n}\,\frac{c^{n}}{\hbar^{n}} ~.
\end{gather}
Ignoring factors of $\mathcal{O}(1)$, the condition $r_\ts{H} \ll L$ becomes
\begin{gather}
	\left(\frac{M_\ts{BH}}{M_{*}}\right)^{\frac{1}{n+1}} \ll \left(\frac{M_\ts{Pl}}{M_{*}}\right)^\frac{2}{n}
\end{gather}
Thus, $M_\ts{BH}/M_*$ should be large enough to justify a semiclassical calculation, but not so large that the black hole begins to notice the extent $L$ of the extra dimensions. In the text we consider theories with fundamental scale as large as $E_{*} \sim 10^{19}\,\text{eV}$, and limit ourselves to UHE cosmic ray collisions with CM energy no greater than $E_\ts{CM} \sim 10^{20}\,\text{eV}$. In that case, for collisions forming black holes with entropy greater than $S_\ts{min}$, the ratio $r_\ts{H}/L$ always satisfies
\begin{gather}
	\frac{r_\ts{H}}{L} \leq 2\pi \left(1.79 \times 10^{17}\right)^{-\frac{1}{n}}\left(\frac{(16\pi)^{2}}{(n+2)^{2}\omega_{n+2}}\right)^{\frac{1}{n}}\,\left(S_\ts{min}\right)^{-\frac{1}{n}} ~.
\end{gather}
For $S_\ts{min}=10^{2}$, this is of order $10^{-18}$ for $n=1$, and of order $10^{-3}$ for $n=6$. In these cases, the black holes we consider are all much smaller than the size of the extra dimensions and the physics should be well-described by formulas that assume an asymptotically flat spacetime. A quick calculation shows that the size of extra dimensions is also much larger than the fundamental length scale $L \gg \ell_{*}$, so that quantum gravity corrections may safely be neglected.

Thus, for black holes with entropy $S_\ts{BH} \geq 10^{2}$, the process of formation via collision and subsequent evaporation should be well described using semiclassical methods and asymptotically flat-space results for $1 \leq n \leq 6$. The case $n=7$ is borderline, with the conditions described above and the assumptions outlined elsewhere in the paper beginning to break down.

\section{Analytic expression for $\mathcal{N}$}
\label{app:AnalyticalN}
Since we consider cosmic rays with energies below the GZK cut-off at $E_\ts{break}$, the minimum energy of a cosmic ray that can participate in a collision with CM energy above $\lambda E_{*}$ is
\begin{gather}
	E_\ts{min} = \frac{(\lambda E_{*})^2}{4\,E_\ts{break}} ~.
\end{gather}
If we express the fundamental scale as a fraction of the maximum value that we can probe, $ E_{*} = (1-\chi)\,2E_\ts{break}/\lambda$ with $0 \leq \chi < 1$, then both cosmic rays must have energy greater than $E_\ts{ankle}$ when
\begin{gather}
	1-\chi > \sqrt{\frac{E_\ts{ankle}}{E_\ts{break}}} = 0.38 ~.
\end{gather}
In this regime the differential flux in Eq.~\eqref{GeneralN} is described by a single power law, and the integral can be evaluated analytically.

Expressing the particle energies in units of $E_\ts{break}$, the number of black holes with $S_\ts{BH} \geq S_\ts{min}$ formed over the past $T = 10^{10}\,\text{yr}$, in a theory with fundamental scale $E_{*} = (1-\chi)\,2E_\ts{break}/\lambda$, is
\begin{multline}\label{eq:AnalyticalN}
\mathcal{N} = 2.38 \times 10^{5}\,(n+2)^2 \, (S_\ts{min})^2\,\left(\frac{1}{1-\chi}\right)^{\frac{2(n+2)}{n+1}} \\
\int_{(1-\chi)^2}^{1} \bns\bns\bns de' de'' du\,u^{\frac{1}{n+1}}\,(e'e'')^{\frac{1}{n+1}-\gamma_2}\,\Theta\big(e'e''u - (1-\chi)^2\big)	
\end{multline}
where $\gamma_2 = 2.59$ is the spectral index given by Auger for cosmic rays with energies between $E_\ts{ankle}$ and $E_\ts{break}$. Notice that $\mathcal{N}$ grows with the minimum entropy for semiclassical calculations as $(S_\ts{min})^2$. This is due to the fact that as $S_\ts{min}$ goes up, the fundamental scales we probe go down like $1/\lambda$, resulting in a larger cross-section \eqref{eq:BHCS2}. 

The full expression obtained from evaluating the integral in Eq.~\eqref{eq:AnalyticalN} is not especially illuminating, but was used to verify the numerical results presented in section \ref{sec:Bounds}. For $\chi \ll 1$, the regime where $E_{*}$ is extremely close to the maximum value for which we can estimate black hole formation rates, $\mathcal{N}$ is well approximated by
\begin{multline}
	\mathcal{N} \simeq 2.38 \times 10^{5}\,(S_\ts{min})^{2} \, \frac{4(n+2)^{2}}{3} \, \chi^{3} \times \\ \left(1 + \gamma_2\,\chi + \frac{4n+5}{2(n+1)}\,\chi \right) ~.
\end{multline}
For $S_\ts{min} = 10^{2}$ and $E_{*} = 10^{18.8}\,\text{eV}$ (corresponding to $\chi = 0.042$) this approximation gives $\mathcal{N} = 2.55 \times 10^{6}$, which is within about $2\%$ of the result obtained directly from Eq.~\eqref{eq:AnalyticalN}.


\newpage
\begin{thebibliography}{1}
 
 
\bibitem{Kanti:2004nr} 
  P.~Kanti,
  ``Black holes in theories with large extra dimensions: A Review,''
  Int.\ J.\ Mod.\ Phys.\ A {\bf 19}, 4899 (2004)
  doi:10.1142/S0217751X04018324
  hep-ph/0402168.
   
\bibitem{Banks:1999gd} 
  T.~Banks and W.~Fischler,
  ``A Model for high-energy scattering in quantum gravity,''
  hep-th/9906038.
  
\bibitem{Giddings:2001bu} 
  S.~B.~Giddings and S.~D.~Thomas,
  ``High-energy colliders as black hole factories: The End of short distance physics,''
  Phys.\ Rev.\ D {\bf 65}, 056010 (2002)
  doi:10.1103/PhysRevD.65.056010
  [hep-ph/0106219].

\bibitem{Giddings:2008gr} 
  S.~B.~Giddings and M.~L.~Mangano,
  ``Astrophysical implications of hypothetical stable TeV-scale black holes,''
  Phys.\ Rev.\ D {\bf 78}, 035009 (2008)
  doi:10.1103/PhysRevD.78.035009
  arXiv:0806.3381 [hep-ph].
  
    
\bibitem{Patrignani:2016xqp} 
  C.~Patrignani {\it et al.} [Particle Data Group],
  ``Review of Particle Physics,''
  Chin.\ Phys.\ C {\bf 40}, no. 10, 100001 (2016).
  doi:10.1088/1674-1137/40/10/100001

\bibitem{Gregory:2013hja} 
  R.~Gregory, I.~G.~Moss and B.~Withers,
  ``Black holes as bubble nucleation sites,''
  JHEP {\bf 1403}, 081 (2014)
  doi:10.1007/JHEP03(2014)081
  [arXiv:1401.0017 [hep-th]].
    
\bibitem{Burda:2015yfa} 
  P.~Burda, R.~Gregory and I.~Moss,
  ``Vacuum metastability with black holes,''
  JHEP {\bf 1508}, 114 (2015)
  doi:10.1007/JHEP08(2015)114
  arXiv:1503.07331 [hep-th].

\bibitem{Burda:2016mou} 
  P.~Burda, R.~Gregory and I.~Moss,
  ``The fate of the Higgs vacuum,''
  JHEP {\bf 1606}, 025 (2016)
  doi:10.1007/JHEP06(2016)025
  arXiv:1601.02152 [hep-th].
  
\bibitem{Cuspinera:2018woe} 
  L.~Cuspinera, R.~Gregory, K.~Marshall and I.~G.~Moss,
  ``Higgs Vacuum Decay from Particle Collisions?,''
  arXiv:1803.02871 [hep-th].
  	
\bibitem{Buttazzo:2013uya} 
  D.~Buttazzo, G.~Degrassi, P.~P.~Giardino, G.~F.~Giudice, F.~Sala, A.~Salvio and A.~Strumia,
  ``Investigating the near-criticality of the Higgs boson,''
  JHEP {\bf 1312}, 089 (2013)
  doi:10.1007/JHEP12(2013)089
  [arXiv:1307.3536 [hep-ph]].
  

\bibitem{Branchina:2014rva} 
  V.~Branchina, E.~Messina and M.~Sher,
  ``Lifetime of the electroweak vacuum and sensitivity to Planck scale physics,''
  Phys.\ Rev.\ D {\bf 91}, 013003 (2015)
  doi:10.1103/PhysRevD.91.013003
  [arXiv:1408.5302 [hep-ph]].
    
\bibitem{Fairbairn:2014zia} 
  M.~Fairbairn and R.~Hogan,
  ``Electroweak Vacuum Stability in light of BICEP2,''
  Phys.\ Rev.\ Lett.\  {\bf 112}, 201801 (2014)
  doi:10.1103/PhysRevLett.112.201801
  [arXiv:1403.6786 [hep-ph]].

\bibitem{Hook:2014uia} 
  A.~Hook, J.~Kearney, B.~Shakya and K.~M.~Zurek,
  ``Probable or Improbable Universe? Correlating Electroweak Vacuum Instability with the Scale of Inflation,''
  JHEP {\bf 1501}, 061 (2015)
  doi:10.1007/JHEP01(2015)061
  [arXiv:1404.5953 [hep-ph]].


\bibitem{Abreu:2011pj} 
  P.~Abreu {\it et al.} [Pierre Auger Collaboration],
  ``The Pierre Auger Observatory I: The Cosmic Ray Energy Spectrum and Related Measurements,''
  arXiv:1107.4809 [astro-ph.HE].
  
  
\bibitem{Greisen:1966jv}  
  K.~Greisen,
  ``End to the cosmic ray spectrum?,''
  Phys.\ Rev.\ Lett.\  {\bf 16}, 748 (1966).
  doi:10.1103/PhysRevLett.16.748

\bibitem{Zatsepin:1966jv} 
  G.~T.~Zatsepin and V.~A.~Kuzmin,
  ``Upper limit of the spectrum of cosmic rays,''
  JETP Lett.\  {\bf 4}, 78 (1966)
  [Pisma Zh.\ Eksp.\ Teor.\ Fiz.\  {\bf 4}, 114 (1966)].

    	
\bibitem{Hut:1983xa} 
  P.~Hut and M.~J.~Rees,
  ``How stable is our vacuum?,''
  Nature {\bf 302}, 508 (1983).
  doi:10.1038/302508a0


\bibitem{1980ApJ...236L..71C}  
  G.~Cunningham, J.~Lloyd-Evans, A.~M.~T.~Pollock, R.~J.~O.~Reid, \& A.~A.~Watson, 
  ``The energy spectrum and arrival direction distribution of cosmic rays with energies above $10^{19}$ electrovolts,''
  \apj~ {\bf 236}, L71 (1980). 
  
\bibitem{Abraham:2010mj} 
  J.~Abraham {\it et al.} [Pierre Auger Collaboration],
  ``Measurement of the energy spectrum of cosmic rays above $10^{18}$ eV using the Pierre Auger Observatory,''
  Phys.\ Lett.\ B {\bf 685}, 239 (2010)
  doi:10.1016/j.physletb.2010.02.013
  [arXiv:1002.1975 [astro-ph.HE]].

  
\bibitem{Aab:2015bza} 
  A.~Aab {\it et al.} [Pierre Auger Collaboration],
  ``The Pierre Auger Observatory: Contributions to the 34th International Cosmic Ray Conference (ICRC 2015),''
  arXiv:1509.03732 [astro-ph.HE].
 
\bibitem{Aab:2017njo}
  A.~Aab {\it et al.} [Pierre Auger Collaboration],
  ``The Pierre Auger Observatory: Contributions to the 35th International Cosmic Ray Conference (ICRC 2017),''
  arXiv:1708.06592 [astro-ph.HE].
 
  
\bibitem{Argyres:1998qn} 
  P.~C.~Argyres, S.~Dimopoulos and J.~March-Russell,
  ``Black holes and submillimeter dimensions,''
  Phys.\ Lett.\ B {\bf 441}, 96 (1998)
  doi:10.1016/S0370-2693(98)01184-8
  [hep-th/9808138].
  

\bibitem{Kaloper:2007pb} 
  N.~Kaloper and J.~Terning,
  ``How black holes form in high energy collisions,''
  Int.\ J.\ Mod.\ Phys.\ D {\bf 17}, 665 (2008)
  [Gen.\ Rel.\ Grav.\  {\bf 39}, 1525 (2007)]
  doi:10.1142/S0218271808012413, 10.1007/s10714-007-0468-5
  [arXiv:0705.0408 [hep-th]].


\bibitem{Myers:1986un} 
  R.~C.~Myers and M.~J.~Perry,
  ``Black Holes in Higher Dimensional Space-Times,''
  Annals Phys.\  {\bf 172}, 304 (1986).
  doi:10.1016/0003-4916(86)90186-7
  
\bibitem{Agashe:2014kda} 
  K.~A.~Olive {\it et al.} [Particle Data Group],
  ``Review of Particle Physics,''
  Chin.\ Phys.\ C {\bf 38}, 090001 (2014).
  doi:10.1088/1674-1137/38/9/090001


\bibitem{Murata:2014nra} 
  J.~Murata and S.~Tanaka,
  ``A review of short-range gravity experiments in the LHC era,''
  Class.\ Quant.\ Grav.\  {\bf 32}, no. 3, 033001 (2015)
  doi:10.1088/0264-9381/32/3/033001
  [arXiv:1408.3588 [hep-ex]].

\bibitem{Hanhart:2001fx} 
  C.~Hanhart, J.~A.~Pons, D.~R.~Phillips and S.~Reddy,
  ``The Likelihood of GODs' existence: Improving the SN1987a constraint on the size of large compact dimensions,''
  Phys.\ Lett.\ B {\bf 509}, 1 (2001)
  doi:10.1016/S0370-2693(01)00544-5
  [astro-ph/0102063].

\bibitem{Hannestad:2003yd} 
  S.~Hannestad and G.~G.~Raffelt,
  ``Supernova and neutron star limits on large extra dimensions reexamined,''
  Phys.\ Rev.\ D {\bf 67}, 125008 (2003)
  Erratum: [Phys.\ Rev.\ D {\bf 69}, 029901 (2004)]
  doi:10.1103/PhysRevD.69.029901, 10.1103/PhysRevD.67.125008
  [hep-ph/0304029].


\bibitem{Sirunyan:2017jix} 
  A.~M.~Sirunyan {\it et al.} [CMS Collaboration],
  ``Search for new physics in final states with an energetic jet or a hadronically decaying $W$ or $Z$ boson and transverse momentum imbalance at $\sqrt{s}=13\text{ }\text{ }\mathrm{TeV}$,''
  Phys.\ Rev.\ D {\bf 97}, no. 9, 092005 (2018)
  doi:10.1103/PhysRevD.97.092005
  [arXiv:1712.02345 [hep-ex]].
          
\end{thebibliography}
\end{document}